- **Technical Report**

# URI Identity and Web Architecture Revisited

**This version**: http://dfdf.inesc-id.pt/tr/doc/web-arch/20071112

**Latest version**: http://dfdf.inesc-id.pt/tr/web-arch

**Author:**

· Xiaoshu Wang (xiao@kdbio.inesc-id.pt)**Abstract**

This document reexamined the URI's identity issue and the debate regarding the nature of "information resource". By making emphasis on the abstract nature of resource and the role of URI as an interface to the web, this article presented an alternative viewpoint about the architecture of the web that would allow us to objectively and consistently treat all kinds of resources.

Note, any comments are welcome at this blog.**Table of Content**

1. Background
    1.1. A brief recap of httpRange-14
    1.2. What is information resource?
    1.3. A case with HTTP content negotiation
2. The Myth about URI's Identity Crisis
3. Web Architecture Revisited
    3.1. The nature of resource
    3.2. Relationship between resource and representation
    3.3. URI Authority vs. Resource Authority
        3.3.1. URI collision
        3.3.2. Resource collision
4. Miscellaneous Issues
    4.1. Do we need Information resource?
    4.2. Can we use 303-Redirect?
    4.3. Another 2xx code?
5. Reference## 1. Background

In the World Wide Web, or simply the web, Uniform Resource Identifier (URI), as defined in [1], is used to identify resources[2]. There has been a long-standing debate about what is the nature of resources that are identified by URIs. The issue was first raised by Kendall Grant Clark on XML.com[3], where he questioned the then circular definition between *resource* and URI . The topic was later discussed by many people, such as Tim Berners-Lee[4], David Booth[5], and Sandro Hawke[6], to just name a few, and it was eventually raised as an architectural issue coined as

httpRange-14 issue[7]. In June 2005, W3C's Technical Architectural Group (TAG) announced that the issue was resolved[8]. Nevertheless, because the resolution was described with the notion of "information resource", question has since been raised about the nature of information resource. To a large degree, or perhaps as a matter of fact, the httpRange-14 resolution has never settled the dispute about a URI's identity. What it did was simply changing the subject. In here, I would like to explain my personal viewpoint on the subject. The viewpoint is neither necessary new nor necessary correct. Nevertheless, it has offered me a consistent ground to understand the web and, subsequently, to guide the design of my projects.

**1.1. A brief recap of httpRange-14**

A URI is commonly assumed to be used for denoting things of different natures. In [5], David Booth summarized that a URI can be used to conjunctively denote a name, a concept, a web location and a document. Since a URI is by nature a name and the current view of URI [1] has discourage the perception of using URI as a locator, the problem can be reduced to the conflict between identifying the concept expressed by a document and the document itself.

Take the URI "http://www.ihmc.us/users/phayes/PatHayes" as an example. Typing the URI in your browser will get back a web page, which content explicitly asserts that the above URI is used to denote a person - Mr. Pat Hayes. What is at debate is if the above behavior should be sanctioned by the current web architecture, and if so, what would be the URI for the returned web page?

In [6], Sandro Hawke proposed to settle the issue with a minor convention of the URI syntax. He suggested using the plain URI for the web page and the same URI appended with an empty fragment identifier (#) for denoting the topic of the page. The proposal, however, met strong resistance from those who prefer to name resources with slash URIs, for example, see the URIs of Dublin Core Metadata Element and Friend of a Friend (FOAF) Vocabulary. In addition, it is not always clear what is *the* topic of a web page. On the other hand, Tim Berners-Lee evaluated a series of models and concluded that there is no better alternative but to limit the scope of what a http-URI can identify[4]. (This may explain why the issue was named as "httpRange" as opposed to a URI identity issue.) After a few years of debate, TAG eventually came up with a resolution. First, TAG concluded that the web architecture should not place arbitrary constraints on what a URI can identify. Second, they recommended a solution for resolving the topic vs. page issue. The solution, for instance, would recommend Mr. Hayes to create another URI for denoting the web page. Thus, when "http://www.ihmc.us/users/phayes/PatHayes" is dereferenced, the client will be 303-redirected to the newly minted URI to retrieve the web page. With this resolution, the concept of Mr. Pat Hayes and a web page describing him are denoted by two different URIs, the ambiguity of URI identity is removed and everyone can be happy.

But the truth is: not everyone is happy. After the httpRange-14 resolution was announced, people are no less, if not more, confounded by the ambiguity of "information resource" than that of "resource". In the subsequent section, I will use two use cases to illustrate the problem.

**1.2. What is information resource?**

One likely consequence of the httpRange-14 resolution is that we may now be able to construct a legitimate reasoning process *to detect* those URIs that has violated the web architecture. For instance, to check if "http://www.ihmc.us/users/phayes/PatHayes" is wrongly implemented, we can take the following steps. (For the sake of brevity, "pat" is used to substitute the long URIs).

1. *pat responds* 200

2. 200 => *pat* is an *information resource* (httpRange-14)
3. *pat* is a person (asserted in the document)
4. A person is not an *information resource* (?)
5. Information resource is disjoint from non-information resource => *pat* is wrong.

There are several ambiguities in the above reasoning. But, let's just discuss the obvious - what makes the assertion of step (4) true? In other words, what is information resource?

From the current description of the architecture of World Wide Web (AWWW)[2], information resource is defined as those resources whose "essential characteristics can be conveyed in a message." But "essential characteristics" appears to be too vague of a wording that can be applied objectively to the real world. For instance, which of the following things would be categorized as an information resource?

1. A book
2. A clock
3. The clock on the wall of my bedroom
4. A gene
5. The sequence of a gene
6. A software
7. A service
8. A namespace
9. An ontology
10. A language
11. A number
12. A concept, such as Dublin Core's creator

I doubt that anyone can give a definite answer. Hence, unless we can build an ontology that arbitrarily divides any conceivable things in the world into two groups and enforce people to use the classification, there is always the question - "what is an information resource?"

**1.3. A case with HTTP content negotiation**

Now, let's take a step back from answering the above challenging question and use 303-redirect whenever in doubt until all endpoints are absolutely information resources. Let's see if this approach may solve the problem.

An image, such as the one shown in Figure 2, should be a clear cut case of information resource. The image is denoted by "http://dfdf.inesc-id.pt/tr/doc/web-arch/img/fig2" and it is represented in several different ways. There are colored image variants that can be requested *via* negotiating for the content type of "image/jpeg" or "image/svg+xml", a black-and-white one for "image/gif", and two text versions for "text/plain" and "text/html", respectively.

Now, the question is: what does "http://dfdf.inesc-id.pt/tr/doc/web-arch/img/fig2" identity? Obviously, the URI denotes something abstract - an idea or a mental image of mine. It should entail all the variant representations but, nevertheless, identify none. The question is: should we now consider an "image" as a non-information resource? Or more generally, should the nature of a resource be changed by virtue of using HTTP content negotiation[i]? If so, then there will not be any information resource - at least within the http space - because any resource may be subject to content negotiation. If not, then what makes a (mental) image so different from a person that makes the former respond 200 but the latter 303? In other words, what is information resource?

To avoid the question, let's step back one more time and try to use the httpRange-14 approach whenever there is an HTTP content negotiation. Per HTTP specification[9], content-negotiation differs from a 303-redirect in that the variant's URI may not be available from the former but must be in the latter. The issue, then, comes down to one key question. That is whether there is a circumstance under which the individual access to a variant's URI is not desired.

In the design of Data Format Description Framework ( DFDF ), a data object is collectively represented by two documents: a data file and a format document. Typically, the data file is in binary form so that data can be stored and accessed more efficiently. The format document, on the other hand, is written in RDF so that ontologies can be used to describe the byte arrangement of the data file as well as the domain semantics of the data. With such a design, data and format are logically bound but physically independent, allowing a data object to maintain its self-descriptiveness while free on choosing data format. Furthermore, to allow the individual reference and access to parts of the binary data, the binary variant of a DFDF data object is designated with a special MIME type - application/dfdf+octet-stream. This particular content type uses a special syntax to construct its fragment identifiers, which interpretation is dependent upon the knowledge from its sibling format variant (Figure 1a).

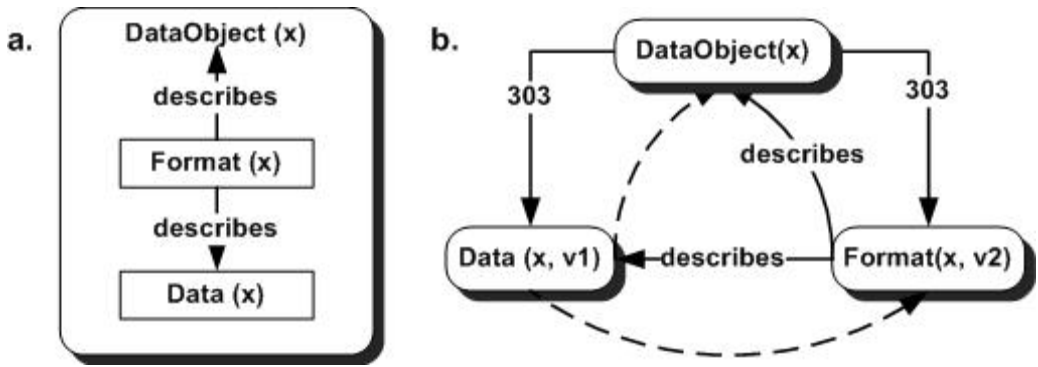

Figure 1 - HTTP Content Negotiation vs. httpRange-14. (a) Binding resources with HTTP content negotiation. (b) Binding resources with 303-redirect. The URI for the entities are enclosed in parenthesis. The arrowed link indicates the possible traverse path from one resource to the other.

But, if the above design is implemented with a 303-approach as shown in Figure 1b, the design will not work properly. The reason is once a variant is converted into a resource with 303, it will be identified by two URIs. This dual name is not a problem *per se* because a resource can have multiple identifiers. What is a problem is that, for a given data object shown in Figure 1b, we cannot assert that "*x#!foo*[ii] *owl:sameAs v1#!foo*" because the interpretation of the former depends on the "*x#foo*" whereas the latter on "*v1#foo*". Because there does not exist a back reference at the 303 endpoint for traversing back to its redirector, it is impossible to locate either *x* or *v2* from *v1* (Figure 1b)[iii].

In summary, within the current architecture of the web, we can neither answer the question - what is information resource - nor avoid it because (1) HTTP content negotiation makes even an obvious case of information resource debatable and (2) we cannot avoid using HTTP content negotiation either because it offers a level of resource encapsulation regarding the use of fragment identifiers that none of other mechanisms can provide. Hence, unless we intend for the web architecture to put arbitrary limit on the application it can support, we are forced to deal with the definition of "information resource".

**2. The Myth about URI's Identity Crisis**

The term *information resource* was invented in large part to resolve the URI's identity crisis. Hence, before diving into the issue of "information resource", we should carefully examine what has really caused the URI identity crisis. To this purpose, let's use another example. Consider, is the following assertion true?

The sixth character of [http://dfdf.inesc-id.pt/tr/doc/web-arch/img/fig2.txt](http://dfdf.inesc-id.pt/tr/doc/web-arch/img/fig2.txt) is 'a'.

Most people would answer 'yes' and in reality this is indeed the case. But, for the sake of argument, let's assume that the sixth character - 'a' - displayed on your browser or text editor is the result of a network error. Now, would you still consider the above assertion correct?

There are two distinct entities involved in this example.

   a. the file parsed by the browser
   b. the file stored on the server

What you see is (a) but what you are asserting is (b). Such an equivocal expression is a common practice in human conversation, where we constantly and subconsciously shift the context of discussion so to refine or even redefine the meaning of a word. Of course, these sorts of equivocations are ambiguous; but most of them would be automatically corrected by our knowledge and experience in the past. On the rare occasions when our past experience has failed, such as the one shown in the above example, the ambiguity will lead to wrong assertions and subsequently miscommunication. The correct expression of the earlier statement should, in fact, be made as follows.

The sixth character of **a representation** of [http://dfdf.inesc-id.pt/tr/doc/web-arch/img/fig1.txt](http://dfdf.inesc-id.pt/tr/doc/web-arch/img/fig1.txt) is 'a'.

This example showed that the identity of URI is never ambiguous. What is ambiguous is our mental assignment of the URI's identity. Similarly, in the Mr. Hayes' example, if we say that "[http://www.ihmc.us/users/phayes/PatHayes](http://www.ihmc.us/users/phayes/PatHayes)" denotes a person and **a representation** of [http://www.ihmc.us/users/phayes/PatHayes](http://www.ihmc.us/users/phayes/PatHayes) is a web page, no confusion would have been created. And in the image example, if we say that [http://dfdf.inesc-id.pt/tr/doc/web-arch/img/fig2](http://dfdf.inesc-id.pt/tr/doc/web-arch/img/fig2) denotes an idea and **one of its representations** is a picture, no ambiguity will arise either.

In essence, the URI's identity crisis is a communication issue as opposed to a technical or a philosophical one. The ambiguity is caused by our attempt to model machine communications to a human one. But, human and machine communications are built on different language systems. The former is psychologically inspired whereas the latter is logic based. A psychologically inspired model suits the human need because it allows us to express unlimited range of thoughts with a manageable set of vocabulary and it has worked wonders in human literature. But such a model will not work for machine, which communication lacks context and continuation so that the meaning of a word must be explicitly specified. For instance, most people would consider the following example as a typical case of URI's ambiguous identity.

<http://www.ihmc.us/users/phayes/PatHayes> a foaf:Person;
    `dc:creator <http://www.ihmc.us/users/phayes/PatHayes>.`

They wonder: what does it mean?

The truth is very simple. It means what it means: the statement asserts that Pat Hayes is a person and the creator of himself. We often *thought* there is ambiguity because we *intended* the above statement to mean that a person created a page as opposed to himself for we *know* that the latter cannot be true. But to reinterpret the subject of dc:creator as a web page in addition to a person has, at the first place, already violated the basic tenet of the web that one URI denotes only one resource. Hence, it should not be a surprise that it leads to a URI's identity crisis. One of the reasons that 303-redirect is recommended by httpRange-14 is to force us decoupling two closely related entities so that we can "unambiguously" express the above semantics as the follows.

```
<http://www.ihmc.us/users/phayes/PatHayes> a foaf:Person.
   <http://www.ihmc.us/users/phayes/PatHayes.html> dc:creator
                    <http://www.ihmc.us/users/phayes/PatHayes>.
```

But what we need, in fact, is simply another vocabulary. For instance, assume the term web:repCreator is defined to refer to an entity that is responsible for the creation of all *representations* of a resource, we can clearly express our intension as the following without creating additional URIs.

<http://www.ihmc.us/users/phayes/PatHayes> a foaf:Person;
      web:repCreator <http://www.ihmc.us/users/phayes/PatHayes>.

Of course, there could be other alternative choices. But the point that I want to raise here is that all the so-called URI identity issue is unwarranted. The URI's ambiguity, if there is one, is caused by our ambiguous wording, which can be simply clarified by using more refined ontological terms. 303-redirect is a solution but, nevertheless, not the only solution.

**3. Web Architecture Revisited**

The architecture of the web is built on three fundamental concepts - URI, resource and representation. A URI is simply a character string adhering to a certain format[1]. It is used to denote a *resource* and dereferencing a URI may get back a *representation*. The relationship of these three entities is illustrated in Figure 2. This depiction differs from the picture shown in the introduction section of the current AWWW document; the change is made to reflect a different viewpoint of the web architecture.

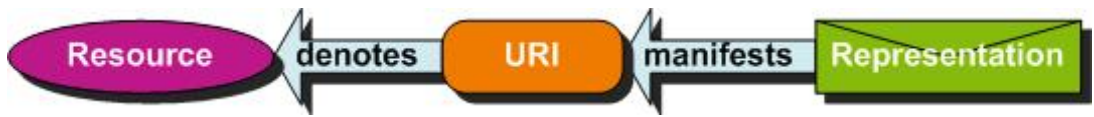

Figure 2 - The relationship between URI, Resource and Representation.

First, the word "*denotes"* is used in place of "*identifies"*. This change is subtle but non-essential; it is made to emphasize the purpose of using URI as a symbol or name so that we will not be misled by the subsense of "identify" as a way to assert the origin or definitive characteristics of a resource. But '*denote*' and '*identify*' can be used interchangeably, as in the many places of this article, as long as the latter is understood in the sense of "establishing an identity" for a resource.

Second, the "represents" relationship between *representation* and *resource* is replaced with a "*manifests*" relationship between *representation* and URI . This change is made to alter the perception that a "representation" is *the equivalent*, or *can take the place*, of a resource, which has contributed for the faulty impression of URI's identity crisis. In the subsequent section, I will discuss how this change may better reflect the architecture of the web.

**3.1. The nature of resource**

Resource was commonly defined by its relationship to either URI or *representation*. The former treats resource as anything that might be identified by a URI [1]; the latter takes resource as a temporal varying relationship to representations. For instance, in the section 5.2.1.1 of his dissertation[10], Roy Fielding writes:

A resource *R* is a temporally varying membership function $M_R(t)$, which for time *t* maps to a set of entities, or values, which are equivalent. The values in the set may be *resource representations* and/or *resource identifiers*….The only thing that is required to be static for a resource is the semantics of the mapping, since the semantics is what distinguishes one resource from another.

But the above description is ambiguous about the nature of the mapped entities. It is unclear if a resource is mapped to the value of a mapped entity or simply the entity itself? To put it more clearly, is a resource mapped to the content of a representation or the representation itself? Or, in the case of a resource identifier, is it mapped to the denoted resource or just the identifier itself. The wording in the last sentence seems to suggest the former because the semantics of mapping is numbered for a given network protocol whereas that of resources is unbounded. Hence unless we assume a resource is mapped to the content of a mapped entity, we could not possibly distinguish all resources from each other.

The content of a representation, however, is not the same as the representation itself. The semantics of the former is about resources whereas that of the latter is about network transportation protocol. Using the above resource definition as it is can easily lead us to ignore the nature of representation as a message of a network protocol and subsequently leads us to believe that a representation can take the place of a resource in the web (See Figure 3).

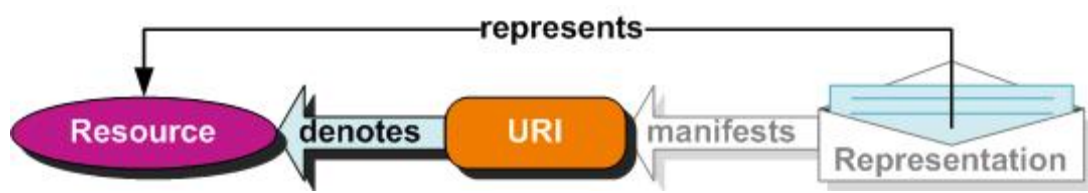

Figure 3 - An easily misled perception about the relationships between resource, URI and representation.

Although the web is commonly described as an information space, in essence, it is a communication system. The interfaces to the web are defined by a set of network protocols, through which we exchange messages about the external world. The web itself, therefore, contains only information about resources but not resources themselves. Resources are connected to the web by binding its URI to a network protocol. The binding is a two part process. For the URI's owner, *representations* of the denoted resource are supplied as the response messages of a network protocol. For the users, the URI is bound as a network request for the representations (See Figure 4). Of course, to ensure a successful communication, the URI's owner and users must agree on binding to the same network protocol. This is commonly achieved through the scheme part of URI but can, nevertheless, be done in other manners. The web architecture itself does not constrain the type of binding because first, URI can be used independently of the web and second, the architecture of the web dictates the orthogonality among specifications.

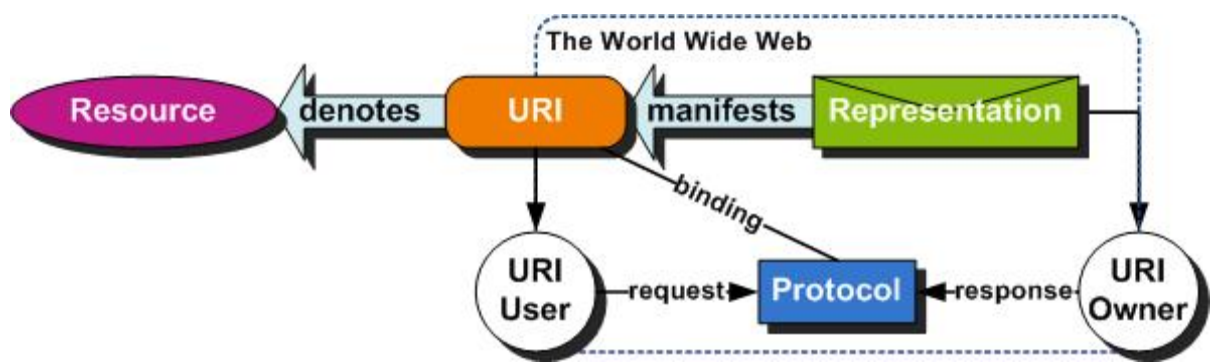

Figure 4 - The architecture of the World Wide Web.

Hence, as far as the web is concerned, resources become abstract entities. Although this conclusion may seem perplexing, the sentiment has been expressed in the past by many people. In the section 6.2.4 of his dissertation, for instance, Roy Fielding wrote that "there are no resources on the server; just mechanisms that supply answers across an abstract interface defined by resources". Similarly, Tim Berners-Lee has also suggested that a generic resource is something like a Platonic ideal [11].

Of course, one may ask: in what sense do we consider *resources* abstract? Here, I would simply treat it as "not being part of the web". The benefit of excluding resources as part of the web is that it can help us to avoid the mental mistake of treating representations as resources. Unlike resources, *representations* are concrete web entities. They are necessary byte-streams and can be directly manipulated within the web. Hence, a person, such as Pat Hayes, does not exist in the web, his representation - a web page - does. A mental image of mine does not exist in the web, its representation - a digital image - does. An electronic text document, no matter how much we think it does, does not exist in the web, its representation - a bit-by-bit copy of the document, does.

Not being resources, however, does not imply that *representations* cannot be the subjects of a message. What can or cannot be described in a message is constrained by the language engaged in the message rather than by the system that delivers the message. In human languages, for instance, we can use phrases, such as "a representation of a URI", to discuss a particular representation. Similarly in RDF, we can do so with a b-node in conjunction with appropriate machine vocabularies. Of course, we can also denote a representation with a canonical URI . But the caveat of this approach is that the URI should not be dereferencible. Denoting a resource with a URI is a different process from binding representations to a URI. The former is a naming/tagging process whereas the latter an abstracting one. Once the URI of a *representation* is bound to a network protocol, the representation disappears from the web but emerged as a *resource*.

By this analysis, *resources* can be further defined as the *abstract entities* that have dereferencible URIs . This not only allows us to forever separate *resources* from *representations* but also allows us to meaningfully combine Roy Fielding's resource definition, which lacks emphasis on the role of URI, with the one in URI specification[1], which lacks the notion of *representation*. Of course, compared to other two definitions, the resource defined in this article has the narrowest scope. As illustrated in Figure 5, there are, in fact, three categories of resources. On the outermost is the resource described by Roy Fielding, whose description reflects a Platonist's account of the web. If we remove the words "resource identifier" from the above definition, the definition can be applied to most, if not any, systems. In other words, the resource defined in Roy Fielding's thesis is very general; it can be anything in the universe. Let's QName this kind resource as "*all:Resource*". Subsumed by *all:Resource* is the resources defined by the URI specification. Because those resources exists in the URI space, let's QName it as *uri:Resource*. At the innermost of the resource-taxonomy is the resource defined in this article. Since they are the resources who are grounded to the web, let's call it "*web:Resource*".

Apparently, the size of *all:Resource* is unbounded and the upper bound of *uri:Resource* of *web:Resources* is the human knowledge. By using this classification, the objective of the web and linked data[12] becomes clear. The former is the effort of mapping human knowledge to machines while the latter making them retrievable within the web so that one day in the future the web as a whole would contain as much information as the human race.

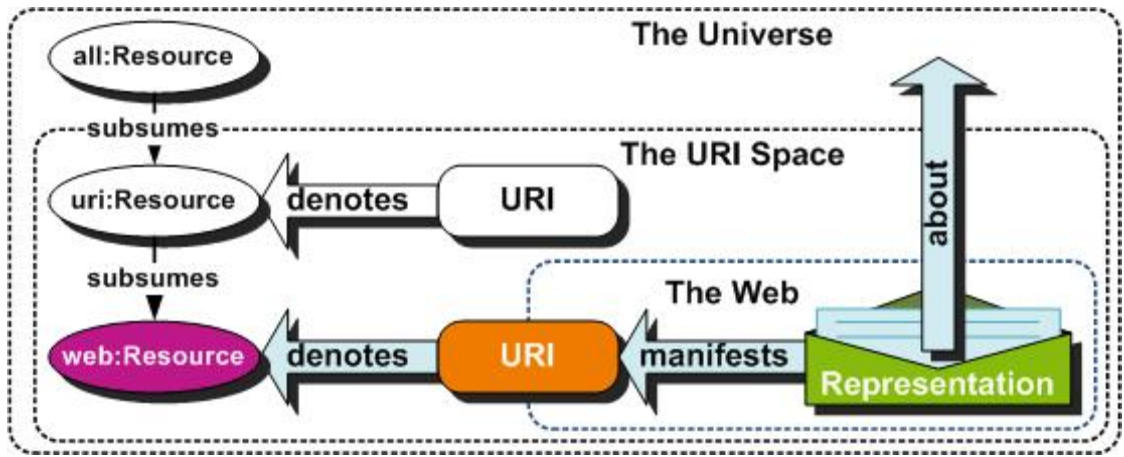

Figure 5: Taxonomy of Resources

**3.2. Relationship between resource and representation**

Although not explicitly documented anywhere, a commonly held belief is that a "representation" bears a somewhat "deterministic" relationship to its denoted resource and all sibling representations are somewhat isomorphic to each other. On example of such perception is the definition of *information resource*[2] as the resources "all of their essential characteristics can be conveyed in a message". But such a perception is incorrectly conceived because it is incompatible with the open world assumption of the web. To illustrate the problem, let's consider an information resource *X* that has two dereferencible URIs - *a* and *b* (See Figure 6). Now, let *C* be the essential characteristics set of *X* and $C_a$, $C_b$ be the set of essential characteristics respectively expressed in $R_a$ and $R_b$, which are the representations of *X* respectively bounded at *a* and *b*. Thus, by the definition of information resource, one of the following two cases must be true:

1. $C_a = C_b = C$
2. $C_a != C$ or $C_b != C$

In the first case, either *a* or *b* is unnecessary implemented. In the second one, either *a* or *b* is a wrongly implemented. But neither case considers the possibilities that $C_a$ and $C_b$ are the proper subset of C. This somehow turns the web into a closed world because the open-world semantics demands only that C entails $C_a$ and/or $C_b$ but not *vice versa*. Hence, unless the web architecture mandates a closed world approach to information resources, a representation cannot be assumed to bear any deterministic or "fully representative" relationship to the denoted resource.

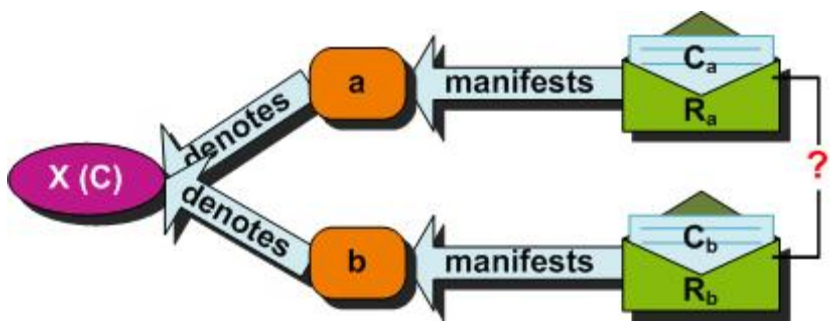

Figure 6 - Relationship between Resource and Representations

By the same token, variant representations delivered via HTTP content negotiation is no exception. Because unless we can assert that all languages and formats share the same expressive power as the others, we must agree that some characteristics of the resource will be lost during the translations. In other words, sibling variants do not necessarily express the same characteristic set of the resource so that they are not necessarily isomorphic to each other.

On a side note, whether a representation should somehow fully represent resource may reflect a different philosophical viewpoint regarding the nature of the web. If we treat the web as an information space, *representations* should be provided to best "represent" the resource. But, on the other hand, if we treat the web as a communicate system, *representations* should be provided to best "communicate" the resource. Personally, I think the latter viewpoint makes more sense and is much easier to work with.

In summary, a web:resource is an abstract entity that has a *representation* in the web; the *representation*, on the other hand, is a byte-stream manifestation of the resource at a particular URI. One resource could have many representations in the web, manifested at one or more URIs, and each representation may or may not reflect the same set of properties of the resource. The architecture of the web does not define how a resource's properties are manifested in a representation[iv] but only how a representation is delivered for a resource. Such a model applies to all web resources regardless if a resource is denoted by a hash- or a slash-URI. The only difference for the former is that the delivery of the representation requires a secondary action from the client.

### 3.3. URI authority vs. resource authority

In addition to the difference between resource and representation, this article would also like to emphasize the distinction between URI authority and resource authority. The lack of distinction has not posted real world problems in the past because most resources were either electronic document or concepts, whose owners are usually the same as, or closely tied to, the URI owners. In addition, as the contents of representation were mostly expressed in natural languages, which do not use URI, there isn't any issue about the inconsistent use of URIs to start with. Nevertheless, as the web is extending into the semantic web, where URIs are routinely used to denote things in the physical world, such as genes, cars, and persons, it should be expected that inconsistent interpretations of URIs and resources will take place. The question is: what role, if any, should the architecture of the web play in case of a conflict?

### 3.3.1. URI collision

URI collision refers to the situation where there are contradicting representations about the resource denoted by the same URI. One such example is shown in Figure 7, where URI "*a*" is allocated to denote a Cat. But elsewhere in the web - "*b*" for instance, *a* is asserted to be a Dog. The question is: what should *a* mean? Please note that the assertion made in "*b*" is not a problem *per se* because, in an open world, *a* can be simply interpreted as a creature that is both a Cat and a Dog. But assuming that we share the common knowledge that Cat is disjoint from Dog, then, the assertion made in "*b*" leads to a contradiction.

Existing AWWW document is ambiguous about the policy for resolving such conflict. The reason is perhaps there is yet an agreement on how to establish the context of interpretation in the semantic web[13]. For instance, OWL uses an explicit import model but RDF uses the follow-your-nose approach (I assume this due to the lack of an import vocabulary in RDF). But, for the sake of brevity, let put this issue aside and assume the context is known. Therefore, returning to the presented example, the question is: within the context of "Dog is disjoint from Cat", what does *a* mean?

a. a Cat
b. a Dog

I think the answer is (a). In other words, I think the representation bound to a URI is more authoritative in defining the meaning of the URI than the representations bound elsewhere. The rational is simple because the alternative disregards URI as the standard interface to the web and makes the whole idea behind the linked data[12] - and the web in general - pointless.

**3.3.2. Resource collision**

Resource collision refers to a different situation from URI collision. In this case, there are contradicting representations about a resource but they are manifested through different URIs. As shown in Figure 7, a dog is denoted by both "*c*" and "*d*". But the representation of "*c*" asserts the dog is smarter than the cat where that of "*d*" asserts just the opposite. Unlike the case of URI collision that begs the question of *which representations about a* URI is more authoritative, the question here is: *which URI is more authoritative in denoting a resource?*

The web, in my opinion, should not answer the above question simply because the web does not directly deal with resources. The web, in this context, is simply a mechanism that helps the URI owners to present their cases by providing representations through the URI. But which URI to choose should be ultimately decided by the web users.

Although the web is built as an open world, in which anyone can say anything about any resource, its monotonic semantics, nevertheless, prevents us from explicitly expressing our discontent about the others. We agree to a representation by making links to its bound URI so to keep the representation alive. We disagree to a representation by refusing to link to it in hopes of the representation will eventually die. Of course, whether a URI will ultimately survive depends on many factors. But the web architecture itself should put as less constrains as possible. The information presented in the web should evolve just as the knowledge of human beings. As Max Planck has puts it, "a new scientific truth does not triumph by convincing its opponents and making them see the light, but rather because its opponents eventually die, and a new generation grows up that is familiar with it [14]."

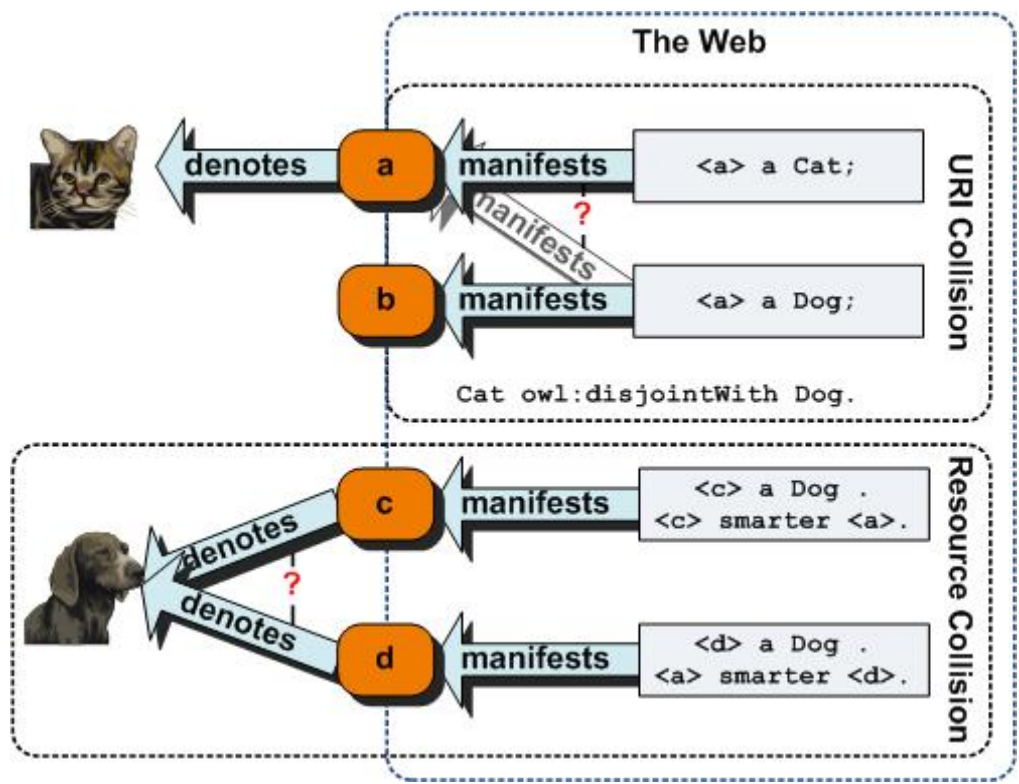

Figure 7 - URI collision vs. Resource Collision

**4. Miscellaneous Issues**

**4.1. Do we need information resource?**

Given the above renewed web architecture, should we still need a definition of "information resource"? I think not because, as Roy Fielding said in this posting to the TAG's mailing list, "doing so solves no known problem in the architecture". But despite Fielding's repeated arguments for his position, for instance, see here and here), I suspect that his viewpoint has not been well received because (1) it is not accurately reflected in the existing AWWW document and (2) it has not been clearly understood by many people, including myself (see how I have implemented the http://proteomicsportal.org), in the past. My position has gradually changed in the past especially during the design of DFDF. The viewpoint presented in this article shares the same viewpoint of Roy Fielding's position but presented the arguments from a slightly different angle.

What motivated the writing of this article is a question posted to W3C's TAG mailing list about if any RDF triples should be drawn from an HTTP interaction, to which I do not think is a good idea because it will break the principle of orthogonal specification that has made the web so successful. To think back, I suspect that the very reason for the question being asked at the first place is the httpRange-14 resolution because it would have enabled the logic inference step suggested in section 1.2. On the surface, the definition of information resource and httpRange-14 may seem harmless. But once they are combined as the ground to ask the above question, they become dangerous to the web architecture. A simple question like "is *rdfs:Resource* an information resource?", for instance, would immediately trap us into the Russell's paradox.

I hope that the presented arguments in this article may help us to conclude the following two points. (1) The definition of "information resource" is not worth debating, at least within the context of the web architecture. (2) What described in httpRange-14 is incorrect. A 200-response code indicates an **informational URI** but not an **information resource**. A resource is not part of the web, hence, it never responds to a request. It is the URI that responds and I hope the revised web architecture (Figure 2, 4) may help to distinguish them.

**4.2. Can we use 303-redirect?**

303-redirect is not architecturally wrong because, just like other HTTP redirects, its semantics is about the location of the message and caching. Hence, of course, it remains as a viable URI design opposite to the straight-forward HTTP content negotiation. There is nothing wrong to use 303 but personally, I prefer content negotiation because it is more efficient, cacheable, and it offers a level of URI hiding that the former cannot provide.

On a side note, according to Roy Fielding's resource definition, the temporal varying relationship to representations is perhaps the most essential characteristics of a resource. Since no one can predict the future yet, the current definition of "information resource" should, in principle, exclude all resources whose representations may contain dynamic or changed content. Should all existing resources be implemented correctly according to httpRange-14, we would have wasted a significant amount of web traffic everyday for solving an issue that we don't even know if it is an issue at the first place.

**4.3. Another 2xx code?**

One of the reasons for using the notion of "information resource" is the comfort of treating web like our desktop so that we can manipulate remote files just as the local ones. Of course, as discussed in this article, this perception can hardly be held in the semantic web. But sometimes we would, nevertheless, still want to use URIs to denote resources, such as a word document, a pdf file or a piece of binary code in the sense that a *representation* is a bit-by-bit copy of the original resource. I think it is reasonable to use a different HTTP response code, for instance, 207 (Bit Copy), to reflect such a relationship without further complications. Of course, the condition of this approach is that the URI cannot be subject to content negotiation because that will make the resource a conceptual object but a binary one.

[i] In this article, the HTTP content negotiation refers to the mechanism described in the HTTP specification, in which the server uses 200 respond code to return a representation. But the idea of content negotiation can also be used with 303, where the server 303 redirects to different URIs according to the Accept- headers.

[ii] The '!' after the '#' is not a typo; it is specifically designed in the specification to indicate the interpretation of the fragment identifier.

[iii] The lack of back-reference is not always a problem. In certain content type, most notably the RDF, any resource can be described. But for most media types, such as image or binary file, it could be a problem because it is not always clear if a resource is a variant of another resource.

[iv] This is the problem set of semantic web as demonstrated by the design of Data Format Description Framework.

Xiaoshu Wang (xiaokdbio.inesc-id.pt)
Last edit: 2007-11-12Z